\documentclass[prd,nofootinbib,twocolumn,superscriptaddress]{revtex4}

\usepackage[colorlinks=true,linkcolor=magenta,urlcolor=blue,citecolor=magenta]{hyperref}

\usepackage{amsfonts,amssymb,amsmath}
\usepackage{epsfig}
\usepackage{mathrsfs}
\usepackage{amssymb}
\usepackage{graphicx}
\graphicspath{ {./images/} }
\usepackage{amsmath}
\usepackage{xcolor}
\usepackage{color}

\begin{document}

\title{Cosmological complexity of the modified dispersion relation}
\author{Tao Li}
\email{2022700450@stu.jsu.edu.cn}
\author{Lei-Hua Liu}
\email{liuleihua8899@hotmail.com}

\affiliation{Department of Physics, College of Physics, Mechanical and Electrical Engineering, Jishou University, Jishou 416000, China}

\begin{abstract}
Complexity will be more and more essential in high-energy physics. It is naturally extended into the very early universe. Considering the universe as a quantum chaotic system, the curvature perturbation of the scalar field is identified with the two-mode squeezed state. By solving the Schr$\ddot{o}$dinger equation, one can obtain the numerical solutions of the angle parameter and squeezing parameter. The solution of the squeezing parameter mainly determines the evolution of complexity. Our numeric indicates that the complexity of the modified dispersion relation will have a non-linear pattern after the horizon exits. Meanwhile, its corresponding Lyapunov index is also larger compared with the standard case. During the inflationary period, the complexity will irregularly oscillate and its scrambling time is also shorter compared with the standard case. Since the modified dispersion relation can be dubbed as the consequences of various frameworks of quantum gravity, it could be applicable to these frameworks. Finally, one can expect the framework of quantum gravity will lead to the fruitful evolution of complexity, which guides us in distinguishing various inflationary models.

\end{abstract}

\maketitle{}

\section{Introduction}
\label{introduction}
The finding of anti-de Sitter/conformal field theory (AdS/CFT) has opened a totally new window and direction for exploring the nature of gravity and strong-coupling system in the condensed field \cite{Maldacena:1997re}. In light of this, spacetime can naturally emerge from quantum entanglement \cite{VanRaamsdonk:2010pw}. Moreover, one can even suspect that ER (Einstein-Rosen bridge) = EPR (Einstein–Prodolsky–Rosen pair) which relates two distinctive physical concepts \cite{Maldacena:2013xja}. Further development can be found in Refs. \cite{Swingle:2014uza, Laanemets:2022rmn, Bianchi:2012ev, Lashkari:2013koa, Balasubramanian:2013lsa}. Thus, quantum entanglement will be more and more essential in high-energy physics. However, the researcher has discovered that the boundary of quantum field theory (QFT) reaches thermal equilibrium within a short temporal interval, and the evolution of wormhole will take a much longer time compared with QFT \cite{Stanford:2014jda}. In order to solve this issue, the so-called quantum complexity is introduced to describe the evolution of the corresponding wormhole \cite{Hartman:2013qma, Liu:2013iza}. In order to relate this concept to the holographic principle, the CA conjecture was proposed by \cite{Brown:2015bva}, in which the complexity is equaled to the computation of action evaluated on a bulk subregion called Wheeler–DeWitt patch. Afterward, there are many profound results were produced \cite{Faulkner:2013ica, Cai:2016xho, Carmi:2017jqz, Carmi:2016wjl, Reynolds:2016rvl, Chapman:2018dem, Chapman:2018lsv, Reynolds:2017lwq}.

 With the development of complexity, two main methods for investigating the so-called circuit complexity are: (a). Nielsen et al. utilize a geometrical method within the phase space of quantum gates \cite{Nielsen2006AGA,Nielsen2006QuantumCA,Dowling2008TheGO}. (b). One can also use the "Fubini-Study" distance to investigate the circuit complexity \cite{Chapman:2017rqy}. In light of these ones, the complexity can be explicitly obtained by using the wave function \cite{Jefferson:2017sdb, Bhattacharyya:2018bbv, Guo:2018kzl} and covariance matrix \cite{Chapman:2018hou, Hackl:2018ptj, Alves:2018qfv, Camargo:2018eof, Ali:2018aon, Khan:2018rzm}. Even though the definition of complexity is still preliminary, Ref. \cite{Jefferson:2017sdb} already tries to give it in quantum field theory \cite{Jefferson:2017sdb}. Based on it, the Hawking radiation can be interpreted by circuit complexity \cite{Zhao:2019nxk, Brown:2019rox}. Futhermore, spacetime can be explained by this complexity \cite{Chandra:2021kdv}. 

In the geometric method \cite{Nielsen2006AGA}, the most convenient quantum system is the inverted Harmonic system. For the single inflationary theory, the Hamilton of curvature perturbation is similar to the inverted Harmonic system in momentum space since the sign of kinetic term and potential is the opposite. Thus, one can implement the technology of quantum circuit complexity to investigate various cosmological periods. Ref.~\cite{Bhargava:2020fhl} has investigated the complexity that is the most essential after the bounce period, in which they also estimated the scrambling time. More generic cosmological models were studied that show the complexity grows fastest in matter-dominated era \cite{Lehners:2020pem}. By the consideration of the backreaction of background expansion \cite{Bhattacharyya:2020rpy}, they found that inflation has the most simple pattern of complexity that will linearly grow after inflation. Ref. \cite{Wang:2022mix} investigated the cosmological complexity of the thermal state in the expansion space. In our previous work \cite{Liu:2021nzx}, we have shown that the formation of primordial blackhole will lead to the different evolution of complexity in inflation, but it is similar after inflation. The complexity is also related to an integral part of the web of diagnostics
for quantum chaos \cite{Roberts:2016hpo, Balasubramanian:2019wgd, Ali:2019zcj, Yang:2019iav, Bhattacharyya:2019txx, Barbon:2019wsy}, which provides the Lyapunov index that could test the Kerr blackhole in various modified gravitational models \cite{Deich:2023oox}. 

Based on Ref. \cite{Liu:2021nzx}, it has shown that the various patterns of complexity in inflation compared with \cite{Bhattacharyya:2020rpy,Bhargava:2020fhl, Lehners:2020pem}, in which they show the complexity in the simple inflationary models. Another point is that our study \cite{Liu:2021nzx} of complexity is only different in inflation. For the late universe, the trend of evolution is similar. In order to capture the various information of complexity in different eras, we will utilize the modified dispersion relation via \cite{Cai:2009hc} since it will impact the power spectrum in various periods due to the modified dispersion relation. From another perspective, the modified dispersion relation can be dubbed as the consequence of quantum gravity \cite{Armendariz-Picon:2003jjq,Armendariz-Picon:2006vgx,Magueijo:2008sx,Piao:2006ja,Martin:2000xs,Arkani-Hamed:2003pdi,Bojowald:2006zb,Jacobson:2000gw,Cai:2007gs}. It has so many phenomenological implications \cite{Cai:2009in,Li:2009rt,Cai:2009zp,Cai:2012yf,Cai:2018tuh,Zheng:2017qfs,Chen:2017dhi,Bianco:2016yib,Pan:2015tza}, $\it i.e.$ the string cosmology, the DBI inflation, the cosmology of loop gravity, $\it e.t.c$. Thus, the investigation of the cosmological complexity of the modified relation is applicable to many theoretical frameworks.

    The structure of this paper is organized as follows. In section \ref{The modified dispersion relation}, we will review the so-called dispersion relation according to \cite{Cai:2009hc}. In section \ref{The squeezed quantum states for cosmological perturbations}, we will treat the cosmological perturbation as the two-mode squeezed state to evaluate the evolution of angle parameter $\phi_k$ and squeezing parameter $r_k$. Section \ref{The complexity of modified dispersion relation} will investigate the evolution of complexity for the modified dispersion relation. In section \ref{summary}, we will give our conclusions and outlooks.

\section{The modified dispersion relation}
\label{The modified dispersion relation}
In this section, we will work in the Friedman–Lemaitre––Robertson Walker background metric ,
\begin{equation}
ds^{2} =a(\eta ) ^{2}(-d\eta ^{2}+d\vec{x} ^{2} ),
\label{frw metric}
\end{equation}
where $\Vec{x}=(x,y,z)$ (denoting the three-dimensional spatial part) is the spatial vector and $a(\eta)$ is the scale factor in conformal time. Since the curvature perturbation is inside the Hubble radius due to the inflationary period, then the wavelength will become larger, meanwhile, the universe undergoes the expansion all the time, and this inhomogeneity will re-enter the Hubble radius again. This physical process can be characterized by which the wavelength $k/a$ is longer than $1/H$. Thus, the conformal time is more convenient compared with physical time. Under metric \eqref{frw metric}, one can define the perturbation of scalar field like
$\phi (x_\mu )=\phi_0(\eta )+\delta \phi (x_\mu )$, the corresponding metric can be read as follows,
\begin{equation}
ds^{2} =a(\eta ) ^{2}\left ( -(1+\psi(\eta,x)  d\eta ^{2}+(1-\psi(\eta,x)d\vec{x}^{2}  \right ) ,
\label{perturbation of metric}
\end{equation}
where \(\psi(\eta,x)\) is the perturbation of metric. Once obtaining metric \eqref{frw metric}, \eqref{perturbation of metric} and the curvature perturbation of scalar field, the perturbated action can be written by  
\begin{equation}
S=\frac{1}{2} \int dtd^3xa^3\frac{\dot{\phi } }{H^2}\left [ \mathcal{\dot{R}}- \frac{1}{a} (\partial _i\mathcal{R} )^2 \right ],
\label{perturbated action}
\end{equation}
where \(H=\frac{\dot{a}}{a}\), \(\mathcal{R}=\psi+\frac{{H} }{\phi _0}\delta\phi\) , $z=\sqrt{2\epsilon}a$, $\epsilon=-\frac{\dot{H}}{H^2}=1-\frac{\mathcal{H}'}{\mathcal{H}^2}$. Action \eqref{perturbated action} can be transferred into a canonical form in terms of the Mukhanov variable \(v =z\mathcal{R}\),
\begin{equation}
S=\frac{1}{2}\int d\eta d^3x\left [ {v}'^2-(\partial _i v )^2+\frac{ {z}'}{z}{v}^2-2\frac{ {z}'}{z}{v}'v  \right ],
\end{equation}
where $\prime$ means the derivative with respect to the conformal time. 
If we approximate $\epsilon$ to be a constant, then one can get $\frac{ {z}'}{z}=\frac{ {a}'}{a}$, 
\begin{equation}
S=\frac{1}{2}\int d\eta d^3x\left [ {v}'^2-(f(k_{\rm ph})\partial _i v)^2+\frac{ {a}'}{a}{v}^2-2\frac{ {a}'}{a}{v}'v  \right ],
\label{starting action} 
\end{equation}
where $f(k_{\rm ph})$ comes the definititon via Ref. \cite{Cai:2009hc},
\begin{equation}
\begin{cases}
  & \text{ if }(k_{\rm ph})>M; f=(\frac{k_{\rm ph}}{M})^{\alpha }  \\
  & \text{ if }(k_{\rm ph})\le M; f=1.
\end{cases}
\label{modified dispersion relation}
\end{equation}
One can see that $f=1$ will nicely recover the standard dispersion relation, which is also equivalent to the sound speed is one as shown in Ref. \cite{Liu:2021nzx}. Furthermore, one could see that $f(k_{\rm ph})$ plays a role of non-trivial sound speed $c_s$ but the origin is different. As discussed in introduction \ref{introduction}, the high energy scale (denoted by $M$) will lead to the modified dispersion relation, such as the trans-Planckian physics and Lorentz violating effects, $\it etc$. However, the formula is different in various theoretical frameworks. As discussed in Refs. \cite{Bianco:2016yib,Lu:2009he}, the modified dispersion relation is all proportional to $k_{\rm ph}^\alpha$ (the notation for describing the modified dispersion is different) where $\alpha$ is non-zero in many gravitational models. That is the reason we adopt the formula $\nu=kf(k_{\rm ph})$ for the dispersion relation. We should emphasize this formula is applicable in many gravitational models since the modified dispersion relation is of the structure $k_{\rm ph}^\alpha$.

Action \eqref{starting action} is our starting point for investigating its evolution of circuit complexity. Varying action with respect to $v$ and then transforming into the momentum space, its equation of motion (EOM) of $v_k$ ($k$ denotes the momentum space) can be derived by
\begin{equation}
{v_k}'' +(\nu^2-\frac{{a}''}{a})v_k=0.
\label{eom of vk}
\end{equation}
$f=1$ corresponds to the EOM of single field inflationary model of curvature perturbation. 

 According to the condition \eqref{modified dispersion relation}, one can clearly see that the dispersion relation can be classified into two regions, one is the so-called high energy scale dubbed as the ultraviolet (UV) regime when $k_{\rm ph}>M$. On the contrary, there is also an infrared (IR) regime as $k_{\rm Ph}<M$. It is naturally considered $M$ as the criterion for assessing the energy scale. Being armed with this logic, the main feature of the dispersion relation can described by the value of $\alpha$, which corresponds to the UV regime or the IR regime. Here, we summarized the values in light of \cite{Cai:2009hc}.
 
 In this kind of relation of the modified dispersion relation, its main feature is characterized by the value of $\alpha$.

$(a)$. The standard dispersion relation corresponds to $\alpha=0$ is also the standard inflationary model \cite{Guth:1980zm}, which belongs to the IR regime

$(b)$. As $0<\alpha<2$, it also belongs to the UV regime.

$(c)$. As $\alpha=2$, it is a special kind of Horv$\check{a}$-Lifshitz cosmology \cite{Kiritsis:2009sh,Calcagni:2009ar}.

$(d)$. As $\alpha>2$, it also belongs to the UV regime.

According to the above understanding, it explicitly shows that theory will be in the UV regime as $\alpha>0$, which is consistent with our previous discussion since the modified dispersion relation comes via the quantum gravitational effects corresponding to $\alpha$ is nonzero. From another aspect, we only assume that $\alpha>0$, since the power spectrum will be conflicted with observation according to $P_{\delta \phi}\propto k^{\alpha}$ as $\alpha<0$. Our discussion only focuses on the power spectrum in light of Ref. \cite{Cai:2009hc}. Therefore, the value of $\alpha$ is independent of specific cosmological models. Furthermore, we will adopt some specific values of $\alpha$ for the following investigations. For intuitively understanding its various cases of $f(k_{\rm ph})$, we give the plot of $f^2$ since it plays the role of non-trivial sound speed. 
\begin{figure}
    \centering
    \includegraphics[width=0.95\linewidth]{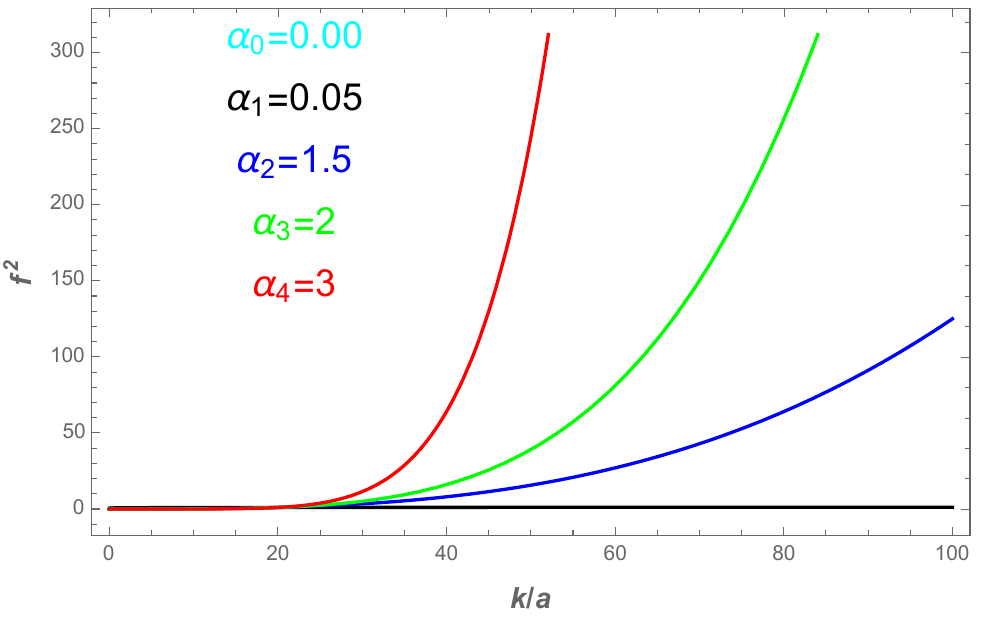}
    \caption{It shows that the modified dispersion relation varies with $k/a$. We have set $M=20$ and set $\alpha=0$ (corresponding to the standard dispersion relation), $\alpha=0.05$, $\alpha=1.5$, $\alpha=2$ (Horv$\check{a}$-Lifshitz gravity) and $\alpha=3$. }
    \label{fig:dispersion}
\end{figure}

Fig. \ref{fig:dispersion} clearly indicates the various cases of $f^2$ varying with comoving momentum. Since it is a simple power-law function in terms of $\alpha$, in which we give five cases: $\alpha=0$, $\alpha=0.05$, $\alpha=1.5$, $\alpha=2$ and $\alpha=3$ that corresponds to our previous illustrations. Here, there is only one thing needs to be noticed that $\alpha_2=1.5$, which we expect that its corresponding complexity could show the deviation compared with the case of the standard dispersion relation. 

\section{The squeezed quantum states for cosmological perturbations}
\label{The squeezed quantum states for cosmological perturbations}

The complexity describes the unstable and chaotic features of a statistical system. The most simple system is the inverted harmonic oscillator, where its corresponding Hamilton is denoted by $H=\frac{1}{2} p^2-\frac{1}{2}kx^2$ with $k$ is the frequency and $p$ is the momentum. Thereafter, one can follow the standard procedure of canonical quantization to investigate its corresponding complexity. Noticing that the sign between the kinetic term and the potential term is opposite, in which the Hamilton of curvature perturbation (can be canonical quantization) has the same situation in momentum space. Thus, it is naturally implemented the technology of quantum information to investigate the complexity of curvature perturbation. Ref. \cite{Albrecht:1992kf} has shown that the curvature perturbation can be transited via the squeezed state, and the wave function of the quantum inverted harmonic oscillator is the Gaussian distribution. Thus, the squeezed state will be implemented into the investigation of complexity for the curvature perturbation of inflation. 

\begin{figure}
    \centering
    \includegraphics[width=1\linewidth]{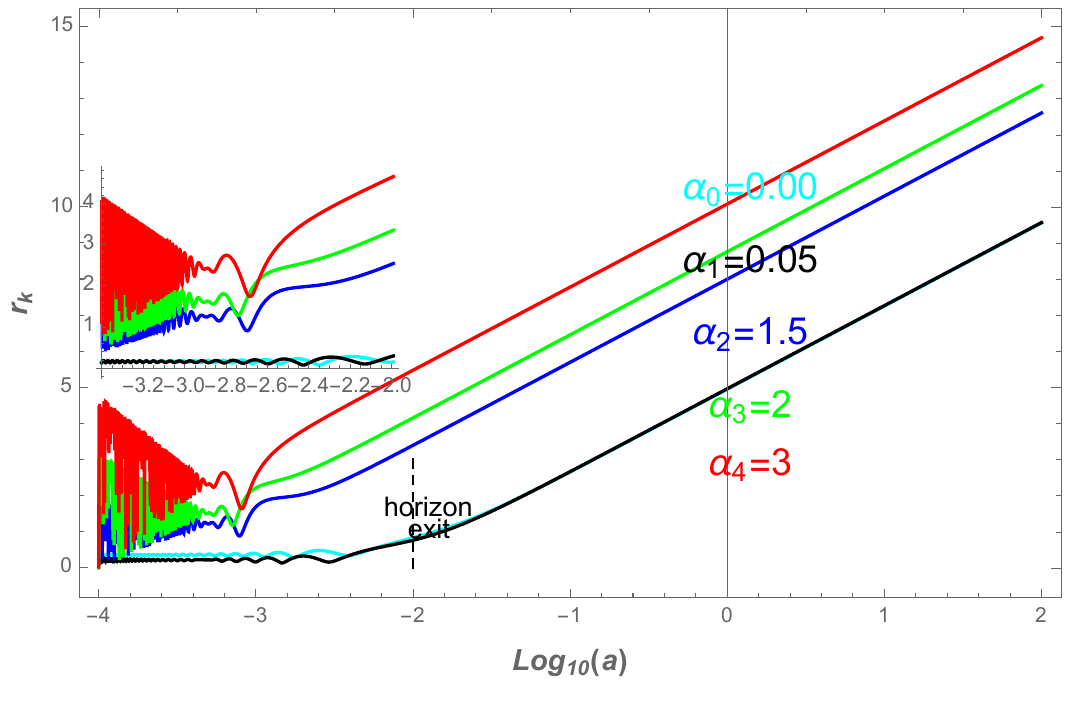}
    \caption{The numerical solutions of $r_k(\eta)$ in terms of \(\log_{10}{a}\) with $ \alpha = 0$,  $ \alpha = 0.05$, $ \alpha = 1.5$, $ \alpha = 2$, and $ \alpha = 3$. Our plots adopt $ H_{0} = 1$.}
    \label{fig:rk}
\end{figure}

\begin{figure}
    \centering
    \includegraphics[width=1\linewidth]{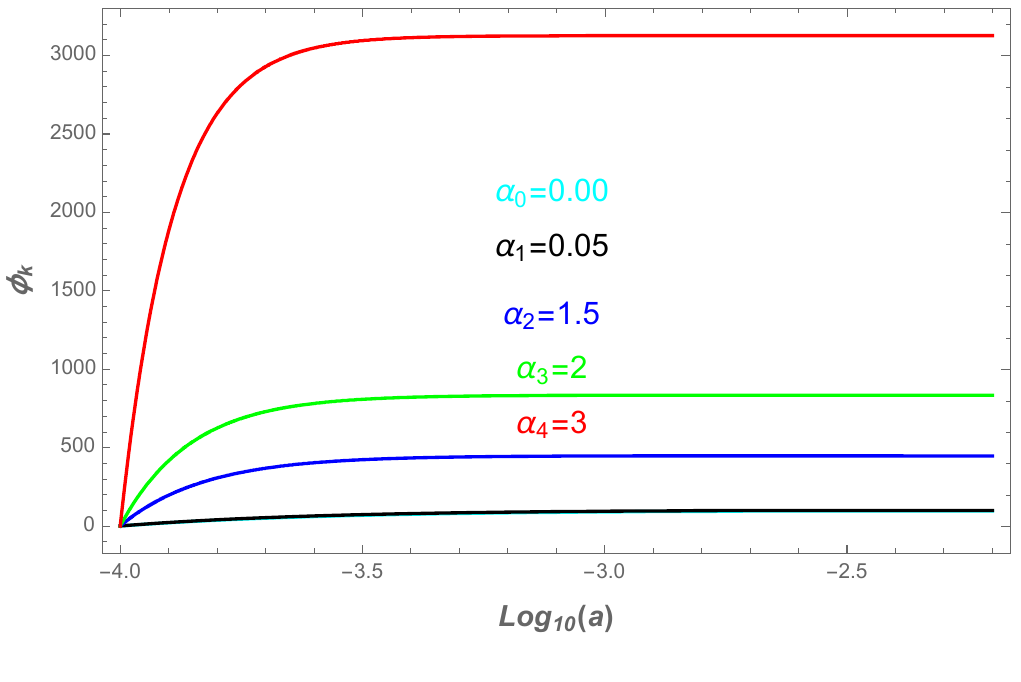}
    \caption{The numerical solutions of $\phi_k(\eta)$ in terms of \(\log_{10}{a}\) with $ \alpha = 0$, $ \alpha = 0.05$, $ \alpha = 1.5$, $ \alpha = 2$, and $ \alpha = 3$. Our plots adopt $ H_{0} = 1$}.
    \label{fig:phik}
\end{figure}

According to EOM \eqref{eom of vk}, one can construct its corresponding action as follows
\begin{equation}
S=\int d\eta L= \frac{1}{2}\int d\eta d^3x\left [ {v}'^2-f^2(\partial _i v)^2+\frac{ {a}'}{a}{v}^2-2\frac{ {a}'}{a}{v}'v  \right ],
\label{action of vi}
\end{equation}
where we do not transform it into the momentum space and $f$ is \eqref{modified dispersion relation}. Once obtaining the action, one can define its canonical momentum,
\begin{equation}
\pi(\eta,\vec{x})=\frac{\delta L}{\delta{v}'(\eta,\vec{x})} ={v}'-\frac{{a}'}{a}v.
\end{equation}
And the Hamiltonian $ H=\int d^3x(\pi v'-\mathcal{L} )$, thus we could obtain
\begin{equation}   
H= \frac{1}{2}\int  d^3x\left [ {\pi}^2+f^2(\partial _i v)^2+\frac{ {a}'}{a}(\pi v+v\pi) \right ].
\end{equation}
Following the standard procedure of quantum field theory, one promote the variable $v$ and $\pi$ as Fourier modes, 
\begin{equation}
\hat{v} (\eta,\vec{x})=\int \frac{d^3k}{(2\pi)^{3/2}} \sqrt{\frac{1}{2k}}(\hat{c }_{-\vec{k} }^{\dagger}v_{\vec {k}}^{\ast }(\eta )e^{-i\vec{k\cdot }\vec{x}}+ {c_{\vec k}}v_{\vec k}e^{i\vec{k\cdot }\vec{x}}),
\end{equation}
\begin{equation}
\hat{\pi} (\eta,\vec{x})=i\int \frac{d^3k}{(2\pi)^{3/2}}\sqrt{\frac{p}{2}}(\hat{c }_{-\vec{k}}^{\dagger}u_{\vec k}^{\ast }(\eta )e^{-i\vec{k\cdot }\vec{x}}-\hat{c}_{\vec k}u_{\vec k}e^{i\vec{k\cdot }\vec{x}}),
\end{equation}
where $\hat{c }_{-\vec{k} }^{\dagger}$ and $\hat{c}_{\vec k}$ represent the creation and annihilation operators, respectively. And then we choose an appropriate normalization condition for mode functions $u_k(\eta)$, $v_k(\eta)$, and we can get the following Hamiltonian
\begin{equation}
\begin{split}
    \hat{H}=\int{d^{3}k}\hat{H}_{k}=&\int{d^{3}k}[\frac{k}{2}(f_{s}^{2}+1)\hat{c }_{-\vec{k} }^{\dagger}\hat{c}_{-\vec{k}}+\frac{k}{2}(f_{s}^{2}+1)\hat{c}_{\vec k}\hat{c}_{\vec k}^{\dagger }\\ &+(\frac{k}{2}(f_{s}^{2}-1)+i\frac{a{}'}{a})\hat{c}_{\vec k}^{\dagger }\hat{c}_{-\vec{k} }^{\dagger }\\ &+(\frac{k}{2}(f_{s}^{2}-1)-i\frac{a{}'}{a})\hat{c}_{\vec k}\hat{c}_{-\vec{k} }].
\end{split}
\label{standard hamilton}
\end{equation}
If $f^2=1$, it will nicely recover the Hamilton of the standard dispersion relation, which is the same as sound speed is one as shown in \cite{Liu:2021nzx}. Observing that Fourier mode contains $\hat{c}_{-\vec{k} }$ and $\hat{c}_{\vec{k}^{\dagger} }$, which the unitary operator should be of two modes. The wave function is Gaussian distribution, one usually implements the unitary evolution operator acting on this Gaussian-type wave function whose form can be parameterized in the factorized form \cite{Albrecht:1992kf,Grishchuk:1990bj} as follows,
\begin{equation}
   \hat{\mathcal{U} }_{\vec{k}}(\eta,\eta_{0} )= \hat{\mathcal{S} }_{\vec{k}}(r_{k},\phi_{k})\hat{\mathcal{R} }_{\vec{k}}(\theta_{k}).
\end{equation}
In the above equation, $\hat{\mathcal{R} }_{\vec{k}}$ is the two-mode rotation operator, which could be written in terms of the rotation angle \(\theta _{k}(\eta)\)
\begin{equation}
   \hat{\mathcal{R} }_{\vec{k}}(\theta _{k})=\exp[-i\theta _{k(\eta)}(\hat{c}_{\vec {k}}\hat{c}_{\vec{k}}^{\dagger }+\hat{c}_{-\vec{ k}}^{\dagger }\hat{c}_{-\vec{ k}})]. 
   \label{rotation operator}
\end{equation}
Meanwhile, $\hat{\mathcal{S} }_{\vec{k}}$ is the two-mode squeeze operator written in terms of the squeezing parameter $r_k(\eta)$ and the squeezing angle $\phi_k(\eta)$, respectively.
\begin{equation}
\hat{\mathcal{S}}_{\vec{k}}(r_{k},\phi_{k})=\exp[r_{k}(\eta)(e^{-2i\phi_{k}(\eta)}\hat{c}_{\vec{k}}\hat{c}_{-\vec{k}}-e^{2i\phi_{k}(\eta)}\hat{c}^{\dagger}_{-\vec{k}}\hat{c}^{\dagger}_{\vec{k}})].  
\label{squeezed operator}
\end{equation}
when $\eqref{rotation operator}$ acts on the initial value of vacuum, it only generates the irrelevant phase factor which will not impact the evolution of wave functions, thus it can be neglected. When using squeezed operator \eqref{squeezed operator} to act on the vacuum \(\left | 0; 0 \right \rangle _{\vec{k},-\vec{k}}\), one can obtain the two-mode squeezed state as follows, 
\begin{equation}
    \left | \psi  \right \rangle_{\rm sq} =\frac{1}{\cosh r_k}\sum_{n=0}^{\infty }(-1)^{n}e^{2in\phi_{k}}\tanh^{n}r_{k} \left | n; n \right \rangle _{\vec{k},-\vec{k}},
    \label{wave function of squeezed state}
\end{equation}
where \( \left | n; n \right \rangle _{\vec{k},-\vec{k}}\) represents the two-mode excited state, which has the following relationship with the two-mode vacuum state \(\left | 0; 0 \right \rangle _{\vec{k},-\vec{k}}\) 
\begin{equation}
    \left | n; n \right \rangle _{\vec{k},-\vec{k}}=\frac{1}{n!}(\hat{c}^{\dagger}_{\vec{k}})^{n}(\hat{c}^{\dagger}_{-\vec{k}})^{n} \left | 0; 0 \right \rangle _{\vec{k},-\vec{k}}.
    \label{excites states}
\end{equation}
Combine Eqs. \eqref{standard hamilton} \eqref{wave function of squeezed state} with Schr\(\ddot{o} \)dinger equation,
\begin{equation}
    i\frac{d}{d\eta} \left | \psi  \right \rangle _{\rm sq}=\hat{H}_{k}\left | \psi  \right \rangle_{\rm sq}
    \label{schrodinger equation},
\end{equation}
when we use the Hamilton operator and $i \frac{d}{d\eta}$ acting on the wave function of squeezed state, the real part and imaginary part would generate the following equations for $\phi_k(\eta)$ and $r_k(\eta)$,
\begin{equation}
 -\frac{dr_{k}}{d\eta}=-\frac{k}{2}(f^{2}-1)\sin(2\phi _{k})+\frac{a{}'}{a}\cos(2\phi _{k}),
 \label{imaginary part}
\end{equation}
\begin{equation}
 \begin{split}
\frac{d\phi_{k}}{d\eta}&=-\frac{k}{2}(f^{2}+1)+\frac{k}{2}(f^{2}-1)\cos(2\phi _{k})\coth(2r_{k})\\ &+\frac{a{}'}{a}\sin(2\phi _{k})\coth(2r_{k}),
\end{split}
\label{real part}
\end{equation}
where Eq. \eqref{imaginary part} comes via the imaginary part of Schr\(\ddot{o} \)dinger equation \eqref{schrodinger equation} and Eq. \eqref{real part} is the real part of Schr\(\ddot{o} \)dinger equation \eqref{schrodinger equation}. To be honest, these two equations are very difficult to solve even for the numerical simulations, thus we change the variable $\eta$ as $\log_{10}a$, and then  Eqs. \eqref{imaginary part}, \eqref{real part} will become as follows,
\begin{equation}
\begin{split}
\frac{10^yH_0}{\rm ln10}\frac{dr}{dy}=\frac{k}{2}[(\frac{k_{\rm ph}}{M})^{2\alpha}-1]\sin(2\phi _{k})-aH_0\cos(2\phi _{k}),
\end{split}
\end{equation}
\begin{equation}
\begin{split}
    \frac{10^yH_0}{\rm ln10}\frac{d\phi_{k}}{dy} =&-\frac{k}{2}[(\frac{k_{\rm ph}}{M})^{2\alpha}+1]+aH_0\sin(2\phi _{k})\coth(2r_{k})\\ &+\frac{k}{2}[(\frac{k_{\rm ph}}{M})^{2\alpha}-1]\cos(2\phi _{k})\coth(2r_{k}),
\end{split}
\end{equation}
where we have defined \(y=\log_{10}{a}\). With these two equations, one numerically simulate $\phi_k$ and $r_k$ in Figs. (\ref{fig:rk},\ref{fig:phik}).  

Fig. \ref{fig:rk}, it shows that the evolution of $r_k$. As $\alpha=0$, it will recover into the standard dispersion relation, also equal to the sound speed $c_s=1$ as shown in \cite{Liu:2021nzx,Bhargava:2020fhl}. In \cite{Liu:2021nzx}, it shows that the non-trivial sound speed will lead, where the deviation cannot be large corresponding to $\alpha<1$, to the fast oscillation and lagging the linear growth. In Fig. \ref{fig:rk}, it clearly indicates the same trend \cite{Liu:2021nzx} when $\alpha<1$ which the $f$ plays a role of $c_s$. The difference comes via the case of $\alpha>1$, in which it shows that there is the damping of the oscillation and the damping will be more significant as enhancing the value of $\alpha$. Another difference is that the time scale for reaching the linear growth after the horizon exit will be shortened as $\alpha>1$, which these cases all belong to the UV regime. Physically speaking, the string theory, trans-Planckian physics, and Lorentz-violating physics will lead to this damping behavior of $r_k$, where it also includes the Horv$\check{a}$-Lifshitz cosmology \cite{Kiritsis:2009sh,Calcagni:2009ar}. In the later investigation of complexity, we will show that the complexity will be mainly influenced by $r_k$. As for $\phi_k$ in Fig. \ref{fig:phik}, the trend of various cases is the same where it will grow until reaching a constant. The only difference is that the constant will be larger as enhancing the value of $\alpha$ which will not significantly impact the complexity. Being armed with these two parameters, one can investigate the evolution of the modified dispersion relation according to Figs. (\ref{fig:rk}, \ref{fig:phik}). Here, we only consider the tree level of the quantum perturbations where we have ignored the contribution from the backreaction of perturbations. However, it may play a significant role from the inflationary scale to the energy scale below self-reproduction \cite{Mukhanov:1996ak}.

\section{The complexity of modified dispersion relation}
\label{The complexity of modified dispersion relation}

Nielson's geometric method will be implemented for investigating the evolution of complexity \cite{Nielsen2006AGA,Nielsen2006QuantumCA,Dowling2008TheGO}. For a given reference state, one can define it as $\left | \psi^{R}  \right \rangle$ at $\tau=0$. At $\tau=1$, the corresponding target state will be connected with the reference state via the unitary operator as follows,
\begin{equation}
   \left | \psi^{T}  \right \rangle_{\tau=1}=U(\tau=1)\left | \psi^{R}  \right \rangle_{\tau=0},
   \label{target state}
\end{equation}
where $\tau$ could parameterize the Hilbert space in quantum mechanics. Following QFT, this unitary operator will be constructed from a path-ordered exponential of a Hamiltonian operator
\begin{equation}
    U(\tau)=\overleftarrow{\mathcal{P}}exp \left (-i \int_{0}^{\tau} dsH(s)  \right ),
    \label{unitory operator}
\end{equation}
where $ \overleftarrow{\mathcal{P}}$ is the path ordering from right to left. In this framework, Hamiltonian operator will be formed in terms of a basis of Hermitian operators $M_{I}$ base, which are the generators for elementary logic gates
\begin{equation}
  H(s)=Y(s)^{I}M_{I},
  \label{hamiltion for gates}
\end{equation}
where $Y(s)^{I}$ is identified with the control function which determines which gate will be switched on of switched off. Meanwhile, it also satisfies with the Schr\(\ddot{o} \)dinger equation
\begin{equation}
    \frac{dU}{ds}= -iY(s)^{I}M_{I}U(s),
    \label{yi}
\end{equation}
In order to define the complexity, one should introduce the function to associated with circuit complexity
 \begin{equation}
     C(U)=\int_{0}^{1}\mathcal{F}(U,\dot{U})d\tau.
     \label{cost function}
 \end{equation}
Then, one can obtain the complexity by minimizing the cost function \eqref{cost function}. Afterword, one can find shortest (geodesic) line between the reference state and target state. First, we will pay attention to the quadratic cost function,
\begin{equation}
    \mathcal{F}(U,Y) =\sqrt{\sum_{l}^{}(Y^{I})^{2}}.
    \label{quadratic cost function}
\end{equation}
The target wave function is the two-mode squeezed state \eqref{wave function of squeezed state}, then one can transform it into the momentum space and we could obtain this,   
\begin{equation}
\begin{split}
    \Psi_{\rm sq}(q_{\vec{{k}}},q_{-\vec{k}} )&= \sum_{n=0}^{\infty }(-1)^{n}\frac{\tanh^{n}r_{k}}{\cosh^{n}r_{k}}\left \langle q_{\vec{k}};q_{-\vec{k}} | n;n  \right \rangle_{\vec{k},-\vec{k}}\\ &=\frac{\exp[A(r_{k},\phi_{k})\cdot(q^{2}_{k}+q^{2}_{-k})-B(r_{k},\phi_{k})q_{\vec{{k}}} q_{-\vec{k} } ]}{\cosh r_{k}\sqrt{\pi}\sqrt{1-e^{-4i\phi_{k}}}\tanh^{2}r_{k}},
\end{split}
\label{wave function1}
\end{equation}
where $A(r_{k},\phi_{k})$ and $B(r_{k},\phi_{k})$ are 
\begin{equation}
    A(r_{k},\phi_{k})=\frac{k}{2}\left (  \frac{e^{-4i\phi_{k}}\tanh^{2}r_{k}+1}{e^{-4i\phi_{k}}\tanh^{2}r_{k}-1} \right ),
    \label{A}
\end{equation}
\begin{equation}
    B(r_{k},\phi_{k})=\frac{k}{2}\left (  \frac{e^{-2i\phi_{k}}\tanh^{2}r_{k}}{e^{-4i\phi_{k}}\tanh^{2}r_{k}-1} \right ).
    \label{B}
\end{equation}
In vector spaces of $ (q_{\vec{{k}}} ,q_{-\vec{k}} )$, Eq. \eqref{wave function1} can be written in terms of diagonal matrix form after some rotation, 
\begin{equation}
\begin{split}
& \Psi_{\rm sq}(q_{\vec{{k}}},q_{-\vec{k}} )=\frac{\exp[-\frac{1}{2}\tilde{M}^{ab}q_{a}q_{b}]}{{\cosh r_{k}\sqrt{\pi}\sqrt{1-e^{-4i\phi_{k}}}\tanh^{2}r_{k}}},
\\ & \tilde{M}= \begin{pmatrix}
 \Omega_{\vec{{k}'} }   & 0\\
  0&\Omega_{{-\vec{{k}'}} }
\end{pmatrix}=\begin{pmatrix}
 -2A+B & 0\\
  0&-2A-B
\end{pmatrix},
\end{split}
\label{the final form of wave function}
\end{equation}
Naturally, one can consider the squeezed vacuum state as the reference state. Then using the same standard procedure to denote it as 
\begin{equation}
\begin{split}
    \Psi_{00}(q_{\vec{{k}}} ,q_{-\vec{k}})&=\left \langle q_{\vec{k}};q_{-\vec{k}} | 0;0  \right \rangle_{\vec{k},-\vec{k}}\\ &=\frac{\exp[-\frac{1}{2}(\omega_{\vec k}q^2_{\vec k}+\omega_{-\vec k}q^2_{-\vec k})] }{\pi^{1/2}}\\ &=\frac{\exp[-\frac{1}{2}\tilde{M}^{ab}q_{a}q_{b}  ]}{\pi^{1/2}}
\end{split}
\label{wave function of vacuum}
\end{equation}
where 
\begin{equation}
\tilde{M}=
    \begin{pmatrix}
 \Omega_{\vec{{k}'} }   & 0\\
  0&\Omega_{{-\vec{{k}'}} }
\end{pmatrix}.
\end{equation}
Note that Eq. \eqref{wave function of vacuum} is the Gaussian distribution that agreed with our previous discussion, and meanwhile the wave function \eqref{the final form of wave function} is also of the Gaussian distribution. Thus, the unitary operator \eqref{unitory operator} will not change the structure of wave functions. According to \eqref{target state}, one can relate the reference states \eqref{wave function of vacuum} \eqref{the final form of wave function} to their corresponding target state via \eqref{unitory operator}, 
\begin{equation}
     \Psi_{\tau}(q_{\vec{{k}}} ,q_{-\vec{k}})=\tilde{U} (\tau)\Psi_{00}(q_{\vec{{k}}} ,q_{-\vec{k}}) \tilde{U}^{\dagger} (\tau),
\end{equation}

\begin{equation}
     \Psi_{\tau=0}(q_{\vec{{k}}} ,q_{-\vec{k}})=\Psi_{00}(q_{\vec{{k}}} ,q_{-\vec{k}})
     \label{boundary condition1},
\end{equation}
\begin{equation}
     \Psi_{\tau=1}(q_{\vec{{k}}} ,q_{-\vec{k}})=\Psi_{\rm sq}(q_{\vec{{k}}} ,q_{-\vec{k}}) 
     \label{boundary condition2},
\end{equation}
where $U (\tau)$ is a $GL (2, C)$ unitary matrix that gives the geodesic line of parameter space between the reference state and target state. Following \cite{Jefferson:2017sdb}, $U (\tau)$ will take the form as follows
\begin{equation}
    \tilde{U} (\tau)=\exp[\sum_{k=1}^{2}Y^{k}(\tau)M^{\rm diag}_{k}],
    \label{tau}
\end{equation}
where ${M^{\rm diag}_{k}}$ denotes two generator of $GL (2, C)$ that defines as 
\begin{equation}   
M^{\rm diag}_{1}=\begin{pmatrix}
 1 & 0\\
  0&0
\end{pmatrix},
   M^{\rm diag}_{2}=\begin{pmatrix}
 0 & 0\\
  0&1
\end{pmatrix}.
\label{generator}
\end{equation}
As we discussed, $U(\tau)$ could parametrize the geodesic line in this group manifold. Thus, the off-diagonal elements are zero since it will generate the nontrivial curvature of this group manifold. As for ${Y_{I} (\tau)}$, it could be constructed from \cite{Ali:2018fcz} as follows, 
\begin{equation}
    Y_{I} (\tau)=Y_{I} (\tau=1)\cdot \tau+Y_{I} (\tau=0).
    \label{Yi}
\end{equation}
From the boundary conditions \eqref{boundary condition1} and \eqref{boundary condition2}, it could obtain that 
\begin{equation}
 \rm Im(Y^{1,2})|_{\tau=0}=\rm Re(Y^{I})|_{\tau=0}=0,
\end{equation}
\begin{equation}
    \rm Im(Y^{1,2})|_{\tau=1}=\frac{1}{2}\ln \frac{|\Omega _{\vec{k},-\vec{k}}|}{\omega _{\vec{k},-\vec{k}}},
\end{equation}
\begin{equation}
  \rm Re(Y^{1,2})|_{\tau=1}=\frac{1}{2}\arctan \frac{\rm Im(\Omega _{\vec{k},-\vec{k}})}{Re(\omega _{\vec{k},-\vec{k}})}. 
\end{equation}
Being armed with these conditions, we can write the complexity as the geodesic line in the parametric manifold as follows, 
\begin{equation}
    C(\tilde{U} )=\int_{0}^{1} d\tau\sqrt{G_{IJ}\dot{Y}^{I}(\tau)\dot{Y}^{I}(\tau)^{*}},
    \label{complexity1}
\end{equation}
where $G_{ij}$ is the induced metric of the group manifold not for the spacetime. In Ref. \cite{Jefferson:2017sdb}, it shows that $G_{IJ}$ could have an arbitrary structure corresponding to various structures of the group manifold. As we mentioned, our group structure is $GL(2, C)$ whose induced metric is flat. Therefore, one can substitute Eq. \eqref{Yi} into Eq. \eqref{complexity1} for obtaining its corresponding complexity,
\begin{equation}
\begin{split}
     C(k)&=\frac{1}{2}[(\ln \frac{|\Omega _{\vec{\rm k}}|}{\omega _{\vec{\rm k}}})^{2}+ (\arctan \frac{\rm {Im}(\Omega _{\vec{k}})}{\rm Re(\omega _{\vec{k}})})^{2}\\ &+(\ln \frac{|\Omega _{-{\vec{\rm k}}}|}{\omega _{-{\vec{\rm k}}}})^{2} +(\arctan \frac{\rm Im(\Omega _{-{\vec{k}}})}{\rm Re(\omega _{-{\vec{k}}})})^{2}], 
\end{split}
\end{equation}
where the information of $\phi_k$ and $r_k$ are including in the coefficients \eqref{A} and \eqref{B}. Then, one can implement the numeric as shown in Figs. (\ref{fig:rk},\ref{fig:phik}) to investigate the evolution of complexity. 
\begin{figure}
    \centering
    \includegraphics[width=1\linewidth]{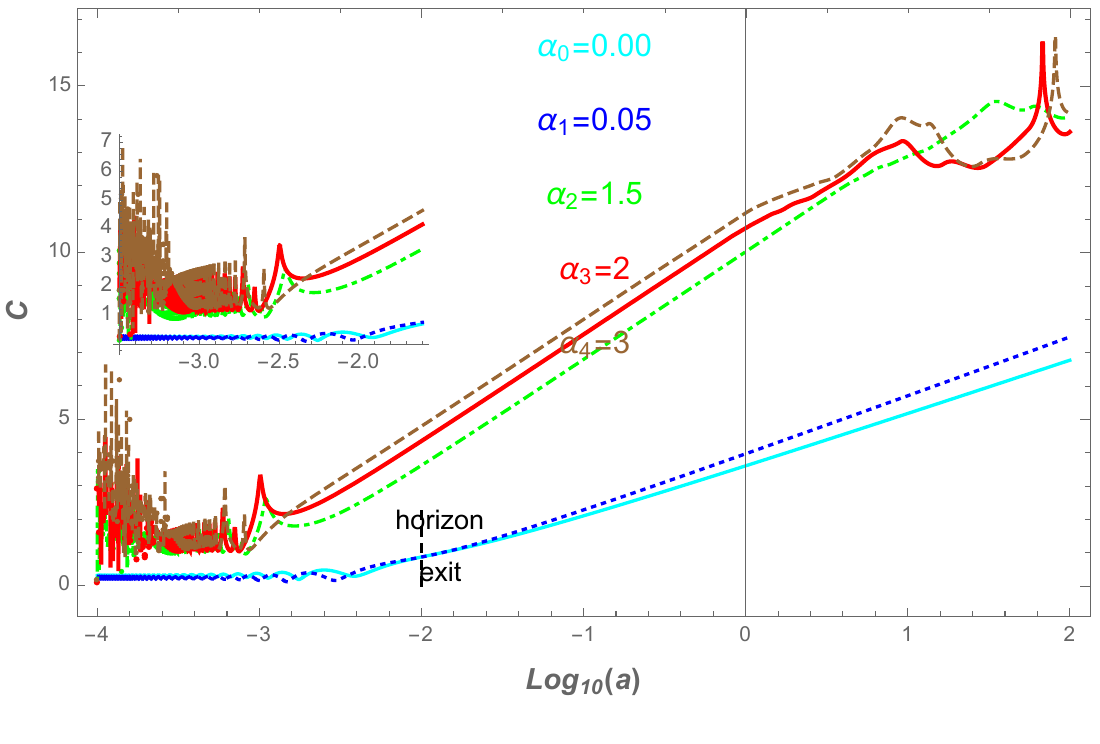}
    \caption{The numerical solutions of $\phi_k(\eta)$ in terms of \(\log_{10}{a}\) with $ \alpha = 0$, $ \alpha = 0.05$, $ \alpha = 1.5$, $ \alpha = 2$, and $ \alpha = 3$. Our plots adopt $ H_{0} = 1$. The scrambling time is identified with the complexity begins to increase and the Lyapunov index can be dubbed as the complexity becomes the linear growth \cite{Ali:2019zcj}.}
    \label{fig:complex11}
\end{figure}

Fig. \ref{fig:complex11} shows the evolution of complexity varying with scale $\ln_{10}(a)$. The most essential physical quantities are scrambling time and the Lyapunov index. In light of Fig. \ref{fig:complex11}, the scrambling time can be defined as the complexity begins to increase and the Lyapunov index can be dubbed as the slope of complexity when complexity becomes linear as shown in \cite{Ali:2019zcj}. In Fig. \ref{fig:complex11}, we can see that the Lyapunov index with $\alpha>1$ is larger compared with $\alpha<1$, which physically means that the framework of string theory, trans-Planckian physics
and Lorentz violating physics will lead to the more chaos comparing with the standard dispersion relation. Our new finding is that the complex will manifest the nonlinear growth as $\alpha>1$ compared with $\alpha<1$ not included in \cite{Liu:2021nzx,Bhattacharyya:2020rpy,Bhargava:2020fhl, Lehners:2020pem}. Thus, we conjecture that if the inflation comes via the string theory, trans-Planckian physics, and Lorentz violating physics, the complexity would show the non-linear evolution after the horizon exit. As for the scrambling time, the cases of $\alpha>1$ will be shorter compared with $\alpha<1$ including the standard dispersion relation. Finally, the oscillation of complexity will occur around the inflationary period, in which one could see that there is no exact set of rules for their damping behaviors as $\alpha>1$ which may be influenced by the evolution of $r_k$ as shown in Fig. \ref{fig:rk}. To sum up, the complexity of $\alpha>1$ will show more chaotic features compared with $\alpha<1$.

\section{Summary and discussion}
\label{summary}
 The modified dispersion relation \eqref{modified dispersion relation} can be considered as the consequences of quantum gravity, including the string cosmology, DBI inflation, cosmology of loop gravity $\it e.t.c.$ As we discussed in Sec. \ref{The modified dispersion relation}, the modified dispersion relation is applicable in many frameworks. From the perspective of quantum information, the dispersion relation could show the various patterns of the evolution of complexity. Based on the above two points, we implement the two-mode squeezed state to investigate the complexity. First and foremost, we write down the Hamilton in terms of creation and annihilation operators \eqref{standard hamilton} in light of the inverted harmonic oscillator. Afterward, we numerically solve \eqref{schrodinger equation} so as to obtain the numeric of $\phi_k$ and $r_k$ as shown in Figs. (\ref{fig:rk},\ref{fig:phik}). Then, we could use them to show how the complexity varies with scale as shown in Fig. \ref{fig:complex11}.   

 The main results of this paper can be summarized as follows: $(a)$. Fig. \ref{fig:rk} clearly indicates that there will be damping behavior as $\alpha>1$, which manifests the effects of the modified dispersion relation. And the oscillation is shorter compared with $\alpha<1$. All of these cases will grow linearly after the horizon exit. As for $\alpha<1$ including the standard dispersion relation, it only shows the oscillation before the horizon exit. $(b)$. Fig. \ref{fig:phik} indicates that the varying trend of $\phi_k$ is the same, where it is growing first and then it approaches a constant value. The only difference is that the maximal value of $\alpha$ will be enhanced by improving $\alpha$. $(c)$. Our new findings are mainly included in Fig. \ref{fig:complex11}. First, it will present the nonlinear evolution after horizon exit in $\alpha>1$, meanwhile this case will also tell the irregular damping oscillations comparing with $\alpha<1$. The slope of complexity when it grows linearly can be identified with the Lyapunov index, which describes the chaotic features of a statistical system. Thus, Fig. \ref{fig:complex11} clearly indicates the Lyapunov index is larger compared with the standard dispersion relation. Another important quantity describing the chaotic system is the scrambling time, which can dubbed as the complexity begins to increase. Obviously, the scrambling time of $\alpha>1$ is shorter compared with $\alpha<1$. Thus, we conjecture that various frameworks of quantum gravity will lead to a more fruitful evolution of the complexity. 

 Our work is based on the single field inflation and the gravitational part only contains the Hilbert-Einstein action, namely $R$ (Ricci scalar). Thus, it naturally extends our methods into $f(R)$ gravitational models \cite{Liu:2018hno, Liu:2018htf} and the multi-field inflationary models \cite{Liu:2020zzv, Liu:2019xhn,Zhang:2022bde,Liu:2021rgq}. The key place is that we need to develop the technology for obtaining their quantum Hamilton in terms of creation and annihilation operators. The k-essence and D-brane naturally contain the modified dispersion relation (equaled to the non-trivial sound speed in some sense) \cite{Li:2021kfq, Li:2021sro}, thus our analysis could also apply to them. Further, we can explicitly implement the modified dispersion relation to the Krylov Complexity \cite{Adhikari:2022whf,Adhikari:2022oxr}.

 \section*{Acknowledgements}
 LH and TL are funded by NSFC grant NO. 12165009 and Hunan Natural Science Foundation NO. 2023JJ30487.

\bibliography{mybibfile}

\section*{References}
\begin{thebibliography}{99}

\bibitem{Maldacena:1997re}
J.~M.~Maldacena,
Adv. Theor. Math. Phys. \textbf{2} (1998), 231-252
doi:10.4310/ATMP.1998.v2.n2.a1
[arXiv:hep-th/9711200 [hep-th]].

\bibitem{VanRaamsdonk:2010pw}
M.~Van Raamsdonk,
Gen. Rel. Grav. \textbf{42} (2010), 2323-2329
doi:10.1142/S0218271810018529
[arXiv:1005.3035 [hep-th]].

\bibitem{Maldacena:2013xja}
J.~Maldacena and L.~Susskind,
Fortsch. Phys. \textbf{61} (2013), 781-811
doi:10.1002/prop.201300020
[arXiv:1306.0533 [hep-th]].

\bibitem{Swingle:2014uza}
B.~Swingle and M.~Van Raamsdonk,
[arXiv:1405.2933 [hep-th]].

\bibitem{Laanemets:2022rmn}
D.~L\"a\"anemets, M.~Hohmann and C.~Pfeifer,
Int. J. Geom. Meth. Mod. Phys. \textbf{19} (2022) no.10, 2250155
doi:10.1142/S0219887822501559
[arXiv:2201.04694 [gr-qc]].

\bibitem{Bianchi:2012ev}
E.~Bianchi and R.~C.~Myers,
Class. Quant. Grav. \textbf{31} (2014), 214002
doi:10.1088/0264-9381/31/21/214002
[arXiv:1212.5183 [hep-th]].

\bibitem{Lashkari:2013koa}
N.~Lashkari, M.~B.~McDermott and M.~Van Raamsdonk,
JHEP \textbf{04} (2014), 195
doi:10.1007/JHEP04(2014)195
[arXiv:1308.3716 [hep-th]].

\bibitem{Balasubramanian:2013lsa}
V.~Balasubramanian, B.~D.~Chowdhury, B.~Czech, J.~de Boer and M.~P.~Heller,
Phys. Rev. D \textbf{89} (2014) no.8, 086004
doi:10.1103/PhysRevD.89.086004
[arXiv:1310.4204 [hep-th]].

\bibitem{Stanford:2014jda}
D.~Stanford and L.~Susskind,
Phys. Rev. D \textbf{90} (2014) no.12, 126007
doi:10.1103/PhysRevD.90.126007
[arXiv:1406.2678 [hep-th]].

\bibitem{Hartman:2013qma}
T.~Hartman and J.~Maldacena,
JHEP \textbf{05} (2013), 014
doi:10.1007/JHEP05(2013)014
[arXiv:1303.1080 [hep-th]].

\bibitem{Liu:2013iza}
H.~Liu and S.~J.~Suh,
Phys. Rev. Lett. \textbf{112} (2014), 011601
doi:10.1103/PhysRevLett.112.011601
[arXiv:1305.7244 [hep-th]].

\bibitem{Brown:2015bva}
A.~R.~Brown, D.~A.~Roberts, L.~Susskind, B.~Swingle and Y.~Zhao,
Phys. Rev. Lett. \textbf{116} (2016) no.19, 191301
doi:10.1103/PhysRevLett.116.191301
[arXiv:1509.07876 [hep-th]].



\bibitem{Faulkner:2013ica}
T.~Faulkner, M.~Guica, T.~Hartman, R.~C.~Myers and M.~Van Raamsdonk,
JHEP \textbf{03} (2014), 051
doi:10.1007/JHEP03(2014)051
[arXiv:1312.7856 [hep-th]].

\bibitem{Cai:2016xho}
R.~G.~Cai, S.~M.~Ruan, S.~J.~Wang, R.~Q.~Yang and R.~H.~Peng,
JHEP \textbf{09} (2016), 161
doi:10.1007/JHEP09(2016)161
[arXiv:1606.08307 [gr-qc]].

\bibitem{Carmi:2017jqz}
D.~Carmi, S.~Chapman, H.~Marrochio, R.~C.~Myers and S.~Sugishita,
JHEP \textbf{11} (2017), 188
doi:10.1007/JHEP11(2017)188
[arXiv:1709.10184 [hep-th]].

\bibitem{Carmi:2016wjl}
D.~Carmi, R.~C.~Myers and P.~Rath,
JHEP \textbf{03} (2017), 118
doi:10.1007/JHEP03(2017)118
[arXiv:1612.00433 [hep-th]].

\bibitem{Reynolds:2016rvl}
A.~Reynolds and S.~F.~Ross,
Class. Quant. Grav. \textbf{34} (2017) no.10, 105004
doi:10.1088/1361-6382/aa6925
[arXiv:1612.05439 [hep-th]].

\bibitem{Chapman:2018dem}
S.~Chapman, H.~Marrochio and R.~C.~Myers,
JHEP \textbf{06} (2018), 046
doi:10.1007/JHEP06(2018)046
[arXiv:1804.07410 [hep-th]].

\bibitem{Chapman:2018lsv}
S.~Chapman, H.~Marrochio and R.~C.~Myers,
JHEP \textbf{06} (2018), 114
doi:10.1007/JHEP06(2018)114
[arXiv:1805.07262 [hep-th]].

\bibitem{Reynolds:2017lwq}
A.~Reynolds and S.~F.~Ross,
Class. Quant. Grav. \textbf{34} (2017) no.17, 175013
doi:10.1088/1361-6382/aa8122
[arXiv:1706.03788 [hep-th]].

	\bibitem{Nielsen2006AGA}
		Nielsen, M.. “A geometric approach to quantum circuit lower bounds.” Quantum Inf. Comput. \textbf{6} (2006): 213-262.

	
		
		\bibitem{Nielsen2006QuantumCA}
	Nielsen, M., M. Dowling, M. Gu and A. Doherty. “Quantum Computation as Geometry.” Science \textbf{311} (2006): 1133 - 1135.
	
	
	

	
	
		\bibitem{Dowling2008TheGO}
	Dowling, M. and M. Nielsen. “The geometry of quantum computation.” Quantum Inf. Comput. \textbf{8} (2008): 861-899.


\bibitem{Chapman:2017rqy}
S.~Chapman, M.~P.~Heller, H.~Marrochio and F.~Pastawski,
Phys. Rev. Lett. \textbf{120} (2018) no.12, 121602
doi:10.1103/PhysRevLett.120.121602
[arXiv:1707.08582 [hep-th]].

\bibitem{Jefferson:2017sdb}
R.~Jefferson and R.~C.~Myers,
JHEP \textbf{10} (2017), 107
doi:10.1007/JHEP10(2017)107
[arXiv:1707.08570 [hep-th]].

\bibitem{Bhattacharyya:2018bbv}
A.~Bhattacharyya, A.~Shekar and A.~Sinha,
JHEP \textbf{10} (2018), 140
doi:10.1007/JHEP10(2018)140
[arXiv:1808.03105 [hep-th]].

\bibitem{Guo:2018kzl}
M.~Guo, J.~Hernandez, R.~C.~Myers and S.~M.~Ruan,
JHEP \textbf{10} (2018), 011
doi:10.1007/JHEP10(2018)011
[arXiv:1807.07677 [hep-th]].

\bibitem{Chapman:2018hou}
S.~Chapman, J.~Eisert, L.~Hackl, M.~P.~Heller, R.~Jefferson, H.~Marrochio and R.~C.~Myers,
SciPost Phys. \textbf{6} (2019) no.3, 034
doi:10.21468/SciPostPhys.6.3.034
[arXiv:1810.05151 [hep-th]].

\bibitem{Hackl:2018ptj}
L.~Hackl and R.~C.~Myers,
JHEP \textbf{07} (2018), 139
doi:10.1007/JHEP07(2018)139
[arXiv:1803.10638 [hep-th]].

\bibitem{Alves:2018qfv}
D.~W.~F.~Alves and G.~Camilo,
JHEP \textbf{06} (2018), 029
doi:10.1007/JHEP06(2018)029
[arXiv:1804.00107 [hep-th]].

\bibitem{Camargo:2018eof}
H.~A.~Camargo, P.~Caputa, D.~Das, M.~P.~Heller and R.~Jefferson,
Phys. Rev. Lett. \textbf{122} (2019) no.8, 081601
doi:10.1103/PhysRevLett.122.081601
[arXiv:1807.07075 [hep-th]].

\bibitem{Ali:2018aon}
T.~Ali, A.~Bhattacharyya, S.~Shajidul Haque, E.~H.~Kim and N.~Moynihan,
Phys. Lett. B \textbf{811} (2020), 135919
doi:10.1016/j.physletb.2020.135919
[arXiv:1811.05985 [hep-th]].

\bibitem{Khan:2018rzm}
R.~Khan, C.~Krishnan and S.~Sharma,
Phys. Rev. D \textbf{98} (2018) no.12, 126001
doi:10.1103/PhysRevD.98.126001
[arXiv:1801.07620 [hep-th]].

\bibitem{Zhao:2019nxk}
Y.~Zhao,
JHEP \textbf{07} (2020), 139
doi:10.1007/JHEP07(2020)139
[arXiv:1912.00909 [hep-th]].

\bibitem{Brown:2019rox}
A.~R.~Brown, H.~Gharibyan, G.~Penington and L.~Susskind,
JHEP \textbf{08} (2020), 121
doi:10.1007/JHEP08(2020)121
[arXiv:1912.00228 [hep-th]].

\bibitem{Chandra:2021kdv}
A.~R.~Chandra, J.~de Boer, M.~Flory, M.~P.~Heller, S.~H\"ortner and A.~Rolph,
JHEP \textbf{21} (2021), 207
doi:10.1007/JHEP04(2021)207
[arXiv:2101.01185 [hep-th]].

\bibitem{Bhargava:2020fhl}
P.~Bhargava, S.~Choudhury, S.~Chowdhury, A.~Mishara, S.~P.~Selvam, S.~Panda and G.~D.~Pasquino,
SciPost Phys. Core \textbf{4} (2021), 026
doi:10.21468/SciPostPhysCore.4.4.026
[arXiv:2009.03893 [hep-th]].

\bibitem{Lehners:2020pem}
J.~L.~Lehners and J.~Quintin,
Phys. Rev. D \textbf{103} (2021) no.6, 063527
doi:10.1103/PhysRevD.103.063527
[arXiv:2012.04911 [hep-th]].

\bibitem{Bhattacharyya:2020rpy}
A.~Bhattacharyya, S.~Das, S.~Shajidul Haque and B.~Underwood,
Phys. Rev. D \textbf{101} (2020) no.10, 106020
doi:10.1103/PhysRevD.101.106020
[arXiv:2001.08664 [hep-th]].

\bibitem{Wang:2022mix}
J.~Wang, H.~Yu and P.~Wu,
[arXiv:2212.01512 [gr-qc]].

\bibitem{Liu:2021nzx}
L.~H.~Liu and A.~C.~Li,
Phys. Dark Univ. \textbf{37} (2022), 101123
doi:10.1016/j.dark.2022.101123
[arXiv:2102.12014 [gr-qc]].

\bibitem{Roberts:2016hpo}
D.~A.~Roberts and B.~Yoshida,
JHEP \textbf{04} (2017), 121
doi:10.1007/JHEP04(2017)121
[arXiv:1610.04903 [quant-ph]].

\bibitem{Balasubramanian:2019wgd}
V.~Balasubramanian, M.~Decross, A.~Kar and O.~Parrikar,
JHEP \textbf{01} (2020), 134
doi:10.1007/JHEP01(2020)134
[arXiv:1905.05765 [hep-th]].

\bibitem{Ali:2019zcj}
T.~Ali, A.~Bhattacharyya, S.~S.~Haque, E.~H.~Kim, N.~Moynihan and J.~Murugan,
Phys. Rev. D \textbf{101} (2020) no.2, 026021
doi:10.1103/PhysRevD.101.026021
[arXiv:1905.13534 [hep-th]].

\bibitem{Yang:2019iav}
R.~Q.~Yang and K.~Y.~Kim,
JHEP \textbf{05} (2020), 045
doi:10.1007/JHEP05(2020)045
[arXiv:1906.02052 [hep-th]].

\bibitem{Bhattacharyya:2019txx}
A.~Bhattacharyya, W.~Chemissany, S.~Shajidul Haque and B.~Yan,
Eur. Phys. J. C \textbf{82} (2022) no.1, 87
doi:10.1140/epjc/s10052-022-10035-3
[arXiv:1909.01894 [hep-th]].

\bibitem{Barbon:2019wsy}
J.~L.~F.~Barb\'on, E.~Rabinovici, R.~Shir and R.~Sinha,
JHEP \textbf{10} (2019), 264
doi:10.1007/JHEP10(2019)264
[arXiv:1907.05393 [hep-th]].

\bibitem{Deich:2023oox}
A.~Deich, N.~Yunes and C.~Gammie,
[arXiv:2308.07232 [gr-qc]].

\bibitem{Cai:2009hc}
Y.~F.~Cai and X.~Zhang,
Phys. Rev. D \textbf{80} (2009), 043520
doi:10.1103/PhysRevD.80.043520
[arXiv:0906.3341 [astro-ph.CO]].

\bibitem{Armendariz-Picon:2003jjq}
C.~Armendariz-Picon and E.~A.~Lim,
JCAP \textbf{12} (2003), 002
doi:10.1088/1475-7516/2003/12/002
[arXiv:astro-ph/0307101 [astro-ph]].

\bibitem{Armendariz-Picon:2006vgx}
C.~Armendariz-Picon,
JCAP \textbf{10} (2006), 010
doi:10.1088/1475-7516/2006/10/010
[arXiv:astro-ph/0606168 [astro-ph]].

\bibitem{Magueijo:2008sx}
J.~Magueijo,
Phys. Rev. D \textbf{79} (2009), 043525
doi:10.1103/PhysRevD.79.043525
[arXiv:0807.1689 [gr-qc]].

\bibitem{Piao:2006ja}
Y.~S.~Piao,
Phys. Rev. D \textbf{75} (2007), 063517
doi:10.1103/PhysRevD.75.063517
[arXiv:gr-qc/0609071 [gr-qc]].

\bibitem{Martin:2000xs}
J.~Martin and R.~H.~Brandenberger,
Phys. Rev. D \textbf{63} (2001), 123501
doi:10.1103/PhysRevD.63.123501
[arXiv:hep-th/0005209 [hep-th]].

\bibitem{Arkani-Hamed:2003pdi}
N.~Arkani-Hamed, H.~C.~Cheng, M.~A.~Luty and S.~Mukohyama,
JHEP \textbf{05} (2004), 074
doi:10.1088/1126-6708/2004/05/074
[arXiv:hep-th/0312099 [hep-th]].

\bibitem{Bojowald:2006zb}
M.~Bojowald, H.~Hernandez, M.~Kagan, P.~Singh and A.~Skirzewski,
Phys. Rev. Lett. \textbf{98} (2007), 031301
doi:10.1103/PhysRevLett.98.031301
[arXiv:astro-ph/0611685 [astro-ph]].

\bibitem{Jacobson:2000gw}
T.~Jacobson and D.~Mattingly,
Phys. Rev. D \textbf{63} (2001), 041502
doi:10.1103/PhysRevD.63.041502
[arXiv:hep-th/0009052 [hep-th]].

\bibitem{Cai:2007gs}
Y.~f.~Cai, M.~z.~Li, J.~X.~Lu, Y.~S.~Piao, T.~t.~Qiu and X.~m.~Zhang,
Phys. Lett. B \textbf{651} (2007), 1-7
doi:10.1016/j.physletb.2007.05.056
[arXiv:hep-th/0701016 [hep-th]].

\bibitem{Cai:2009in}
Y.~F.~Cai and E.~N.~Saridakis,
JCAP \textbf{10} (2009), 020
doi:10.1088/1475-7516/2009/10/020
[arXiv:0906.1789 [hep-th]].


\bibitem{Li:2009rt}
M.~Li, Y.~F.~Cai, X.~Wang and X.~Zhang,
Phys. Lett. B \textbf{680} (2009), 118-124
doi:10.1016/j.physletb.2009.08.053
[arXiv:0907.5159 [hep-ph]].

\bibitem{Cai:2009zp}
Y.~F.~Cai, E.~N.~Saridakis, M.~R.~Setare and J.~Q.~Xia,
Phys. Rept. \textbf{493} (2010), 1-60
doi:10.1016/j.physrep.2010.04.001
[arXiv:0909.2776 [hep-th]].

\bibitem{Cai:2012yf}
Y.~F.~Cai, M.~Li and X.~Zhang,
Phys. Lett. B \textbf{718} (2012), 248-254
doi:10.1016/j.physletb.2012.10.065
[arXiv:1209.3437 [hep-th]].

\bibitem{Cai:2018tuh}
Y.~F.~Cai, X.~Tong, D.~G.~Wang and S.~F.~Yan,
Phys. Rev. Lett. \textbf{121} (2018) no.8, 081306
doi:10.1103/PhysRevLett.121.081306
[arXiv:1805.03639 [astro-ph.CO]].

\bibitem{Zheng:2017qfs}
Y.~Zheng, L.~Shen, Y.~Mou and M.~Li,
JCAP \textbf{08} (2017), 040
doi:10.1088/1475-7516/2017/08/040
[arXiv:1704.06834 [gr-qc]].

\bibitem{Chen:2017dhi}
J.~Chen, W.~Hou, D.~Hou and T.~Qiu,
Chin. Phys. C \textbf{42} (2018) no.4, 045102
doi:10.1088/1674-1137/42/4/045102
[arXiv:1711.06580 [astro-ph.CO]].

\bibitem{Bianco:2016yib}
S.~Bianco, V.~N.~Friedhoff and E.~Wilson-Ewing,
Phys. Rev. D \textbf{97} (2018) no.4, 046006
doi:10.1103/PhysRevD.97.046006
[arXiv:1609.06891 [gr-qc]].

\bibitem{Pan:2015tza}
W.~J.~Pan and Y.~C.~Huang,
Gen. Rel. Grav. \textbf{48} (2016) no.11, 144
doi:10.1007/s10714-016-2138-y
[arXiv:1508.06475 [hep-th]].

\bibitem{Lu:2009he}
Y.~Lu and Y.~S.~Piao,
Int. J. Mod. Phys. D \textbf{19} (2010), 1905-1914
doi:10.1142/S0218271810018074
[arXiv:0907.3982 [hep-th]].

\bibitem{Guth:1980zm}
A.~H.~Guth,
Phys. Rev. D \textbf{23} (1981), 347-356
doi:10.1103/PhysRevD.23.347

\bibitem{Kiritsis:2009sh}
E.~Kiritsis and G.~Kofinas,
Nucl. Phys. B \textbf{821} (2009), 467-480
doi:10.1016/j.nuclphysb.2009.05.005
[arXiv:0904.1334 [hep-th]].

\bibitem{Calcagni:2009ar}
G.~Calcagni,
JHEP \textbf{09} (2009), 112
doi:10.1088/1126-6708/2009/09/112
[arXiv:0904.0829 [hep-th]].

\bibitem{Albrecht:1992kf}
A.~Albrecht, P.~Ferreira, M.~Joyce and T.~Prokopec,
Phys. Rev. D \textbf{50} (1994), 4807-4820
doi:10.1103/PhysRevD.50.4807
[arXiv:astro-ph/9303001 [astro-ph]].

\bibitem{Grishchuk:1990bj}
L.~P.~Grishchuk and Y.~V.~Sidorov,
Phys. Rev. D \textbf{42} (1990), 3413-3421
doi:10.1103/PhysRevD.42.3413

\bibitem{Mukhanov:1996ak}
V.~F.~Mukhanov, L.~R.~W.~Abramo and R.~H.~Brandenberger,
Phys. Rev. Lett. \textbf{78} (1997), 1624-1627
doi:10.1103/PhysRevLett.78.1624
[arXiv:gr-qc/9609026 [gr-qc]].






\bibitem{Ali:2018fcz}
T.~Ali, A.~Bhattacharyya, S.~Shajidul Haque, E.~H.~Kim and N.~Moynihan,
JHEP \textbf{04} (2019), 087
doi:10.1007/JHEP04(2019)087
[arXiv:1810.02734 [hep-th]].















\bibitem{Liu:2018hno}
L.~H.~Liu, T.~Prokopec and A.~A.~Starobinsky,
Phys. Rev. D \textbf{98} (2018) no.4, 043505
doi:10.1103/PhysRevD.98.043505
[arXiv:1806.05407 [gr-qc]].

\bibitem{Liu:2018htf}
L.~H.~Liu,
doi:10.1007/s10773-018-3809-0
[arXiv:1807.00666 [gr-qc]].

\bibitem{Liu:2020zzv}
L.~H.~Liu and T.~Prokopec,
JCAP \textbf{06} (2021), 033
doi:10.1088/1475-7516/2021/06/033
[arXiv:2005.11069 [astro-ph.CO]].

\bibitem{Liu:2019xhn}
L.~H.~Liu and W.~L.~Xu,
Chin. Phys. C \textbf{44} (2020) no.8, 085103
doi:10.1088/1674-1137/44/8/085103
[arXiv:1911.10542 [astro-ph.CO]].

\bibitem{Zhang:2022bde}
X.~z.~Zhang, L.~h.~Liu and T.~Qiu,
Phys. Rev. D \textbf{107} (2023) no.4, 043510
doi:10.1103/PhysRevD.107.043510
[arXiv:2207.07873 [hep-th]].

\bibitem{Liu:2021rgq}
L.~H.~Liu,
Chin. Phys. C \textbf{47} (2023), 1
doi:10.1088/1674-1137/ac9d28
[arXiv:2107.07310 [astro-ph.CO]].



\bibitem{Li:2021kfq}
A.~c.~Li, X.~F.~Li, D.~f.~Zeng and L.~H.~Liu,
[arXiv:2102.12939 [gr-qc]].

\bibitem{Li:2021sro}
A.~c.~Li,
[arXiv:2110.15855 [hep-th]].







\bibitem{Adhikari:2022whf}
K.~Adhikari, S.~Choudhury and A.~Roy,
Nucl. Phys. B \textbf{993} (2023), 116263
doi:10.1016/j.nuclphysb.2023.116263
[arXiv:2204.02250 [hep-th]].


\bibitem{Adhikari:2022oxr}
K.~Adhikari and S.~Choudhury,
Fortsch. Phys. \textbf{70} (2022) no.12, 2200126
doi:10.1002/prop.202200126
[arXiv:2203.14330 [hep-th]].












\end{thebibliography}
\end{document}